\documentclass[twocolumn,twocolumn]{IEEEtran}
\usepackage[T1]{fontenc}
\usepackage{color}
\usepackage{float}
\usepackage{mathrsfs}
\usepackage{amsmath}
\usepackage{amssymb}
\usepackage{graphicx}
\usepackage{amsthm} 

\usepackage[unicode=true,
bookmarks=true,bookmarksnumbered=true,bookmarksopen=true,bookmarksopenlevel=1,
breaklinks=false,pdfborder={0 0 0},pdfborderstyle={},backref=false,colorlinks=false]
{hyperref}
\hypersetup{pdftitle={Your Title},
	pdfauthor={Your Name},
	pdfpagelayout=OneColumn, pdfnewwindow=true, pdfstartview=XYZ, plainpages=false}

\makeatletter

\floatstyle{ruled}
\newfloat{algorithm}{tbp}{loa}
\providecommand{\algorithmname}{Algorithm}
\floatname{algorithm}{\protect\algorithmname}

\usepackage[caption=false,font=footnotesize]{subfig}
\usepackage{algorithm}
\usepackage{algorithmic}

\usepackage{multirow} 
\usepackage{amsmath} 
\usepackage{xcolor}

\allowdisplaybreaks[4]

\ifCLASSOPTIONcompsoc
\usepackage[caption=false,font=normalsize,labelfont=sf,textfont=sf]{subfig}
\else
\usepackage[caption=false,font=footnotesize]{subfig}
\fi

\usepackage{cite}
\usepackage{bm}
\usepackage{algorithmic}
\usepackage{algorithm}
\usepackage{graphicx}
\IEEEoverridecommandlockouts
\usepackage{lettrine}

\usepackage{geometry}
\@ifundefined{showcaptionsetup}{}{%
	\PassOptionsToPackage{caption=false}{subfig}}
\usepackage{subfig}
\makeatother
\begin{document}

\title{\textcolor{black}{Robust Task Offloading for UAV-enabled Secure MEC Against Aerial Eavesdropper}}
\author{
	\IEEEauthorblockN{Can Cui$^{\dagger}$, Ziye Jia$^{\dagger\ast }$, Chao Dong$^{\dagger}$ and Qihui Wu$^{\dagger}$\\
	\IEEEauthorblockA{$^{\dagger}$The Key Laboratory of Dynamic Cognitive System of Electromagnetic Spectrum Space, Ministry of Industry and Information Technology, Nanjing University of Aeronautics and Astronautics, Nanjing, Jiangsu, 211106, China\\
	$^{\ast }$National Mobile Communications Research Laboratory, Southeast University, Nanjing, Jiangsu, 211111, China\\
	\{cuican020619, jiaziye, dch, wuqihui\}@nuaa.edu.cn}
	\thanks{This work was supported by the National Key R{\&}D Program of China under Grant 2022YFB3104502, in part by National Natural Science Foundation of China under Grant 62301251, in part by  the open research fund of National Mobile Communications Research Laboratory, Southeast University (No. 2024D04), in part by the Aeronautical Science Foundation of China 2023Z071052007, and in part by the Young Elite Scientists Sponsorship Program by CAST 2023QNRC001.}
	}
}

\maketitle
\pagestyle{empty} 

\thispagestyle{empty}
\begin{abstract}
Unmanned aerial vehicles (UAVs) are recognized as a promising candidate for the multi-access edge computing (MEC) in the future sixth generation communication networks. However, the aerial eavesdropping UAVs (EUAVs) pose a significant security threat to the data offloading. In this paper, we investigate a robust MEC scenario with multiple service UAVs (SUAVs) towards the potential eavesdropping from the EUAV, in which the random parameters such as task complexities are considered in the practical applications. In detail, the problem is formulated to optimize the deployment positions of SUAVs, the connection relationships between GUs and SUAVs, and the offloading ratios. With the uncertain task complexities, the corresponding chance constraints are constructed under the uncertainty set, which is tricky to deal with. Therefore, we first optimize the pre-deployment of SUAVs by the K-means algorithm. Then, the distributionally robust optimization method is employed, and the conditional value at risk is utilized to transform the chance constraints into convex forms, which can be solved via convex toolkits. Finally, the simulation results show that with the consideration of uncertainties, just 5\% more energy is consumed compared with the ideal circumstance, which verifies the robustness of the proposed algorithms.
\end{abstract}


\section{Introduction\label{sec1}}
\lettrine[lines=2]{T}{he} multi-access edge computing (MEC) has emerged as an effective solution to the computation-intensive tasks in the future sixth generation of communication system. However, it is unaffordable to realize the ubiquitous MEC services merely through deploying ground infrastructures. To deal with this challenge, the unmanned aerial vehicle (UAV), which is a promising candidate for MEC, allows the ground users (GUs) to offload their tasks by integrating edge servers \cite{Computation-You}. Alternatively, the deployment of UAVs enables a wider range of applications and provides more sufficient computation services in the areas lacking infrastructures compared with the ground networks \cite{Joint-Lu,Service-Jia,Cooperative-Jia}. Nevertheless, the existence of potential aerial eavesdroppers brings a threat for the multiple UAVs enabled MEC paradigm. Since the malicious UAVs intercept the task information and overhear the communication link, it is necessary to focus on the secure offloading issue especially in the open airspace \cite{DDQN-Ding}.

A couple of works have been investigated to resist eavesdropping and enhance the reliability of the UAV-enabled MEC network. For instance, \cite{Against-Zhao} focused on the security challenges in UAV-assisted MEC networks to maximize the secrecy transmission rate. \cite{Task-Zhang} designed a joint dynamic programming and bidding algorithm to improve the security performance for the UAVs assisted MEC scenario. \cite{Collaborative-Ding} focused on the collaborative communication and computation against aerial eavesdropping via the block coordinate descent and deep reinforcement learning methods. Considering the security threat from malicious UAVs, \cite{Green-Zhao} designed an optimization framework for secure offloading based on the deep reinforcement learning. However, in practical applications, the computation complexities of tasks are random, and their exact distributions are difficult to obtain. These uncertainties may result in unexpected computation outages, which are ignored in most available researches.

In response to the above issues, in this paper, we highlight the security issue in the aerial MEC scenario, with multiple service UAVs (SUAVs) equipped with edge servers and an eavesdropping UAV (EUAV) monitoring the data links. We aim to minimize the total energy consumption by optimizing the deployment positions of SUAVs, the connection relationships of GU-SUAV, and the offloading ratios. The secure offloading problem is formulated with chance constraints under the uncertainty set for random task complexities, which is in the form of mixed integer nonlinear programming. Besides, the lack of the distribution information of the uncertain task complexities leads to the difficulty in problem solving. To deal with the proposed problem, the K-means based algorithm is developed for the SUAV deployment. Then, with the determined positions of SUAVs and connection relationships, we utilize the distributionally robust optimization (DRO) method and transform the chance constraints into distributionally robust chance constraints (DRCCs). The concept and theory on conditional value at risk (CVaR) are further introduced for the approximate conservative solutions to the problem with DRCCs. Then, the DRCCs are transformed into the convex form of second order cone programming (SOCP) and solved via CVX. Finally, we conduct numerical simulations to verify the robustness of the proposed algorithms. 

The remaining of this paper is organized as follows. The problem is formulated in Section \ref{sec2} and the algorithms are designed in Section \ref{sec3}. Simulation results are provided in Section \ref{sec4}. Finally, we draw conclusions in Section \ref{sec5}.
\section{System Model and Problem Formulation\label{sec2}}

\begin{figure}[t]
	\centering{\includegraphics[scale=0.25]{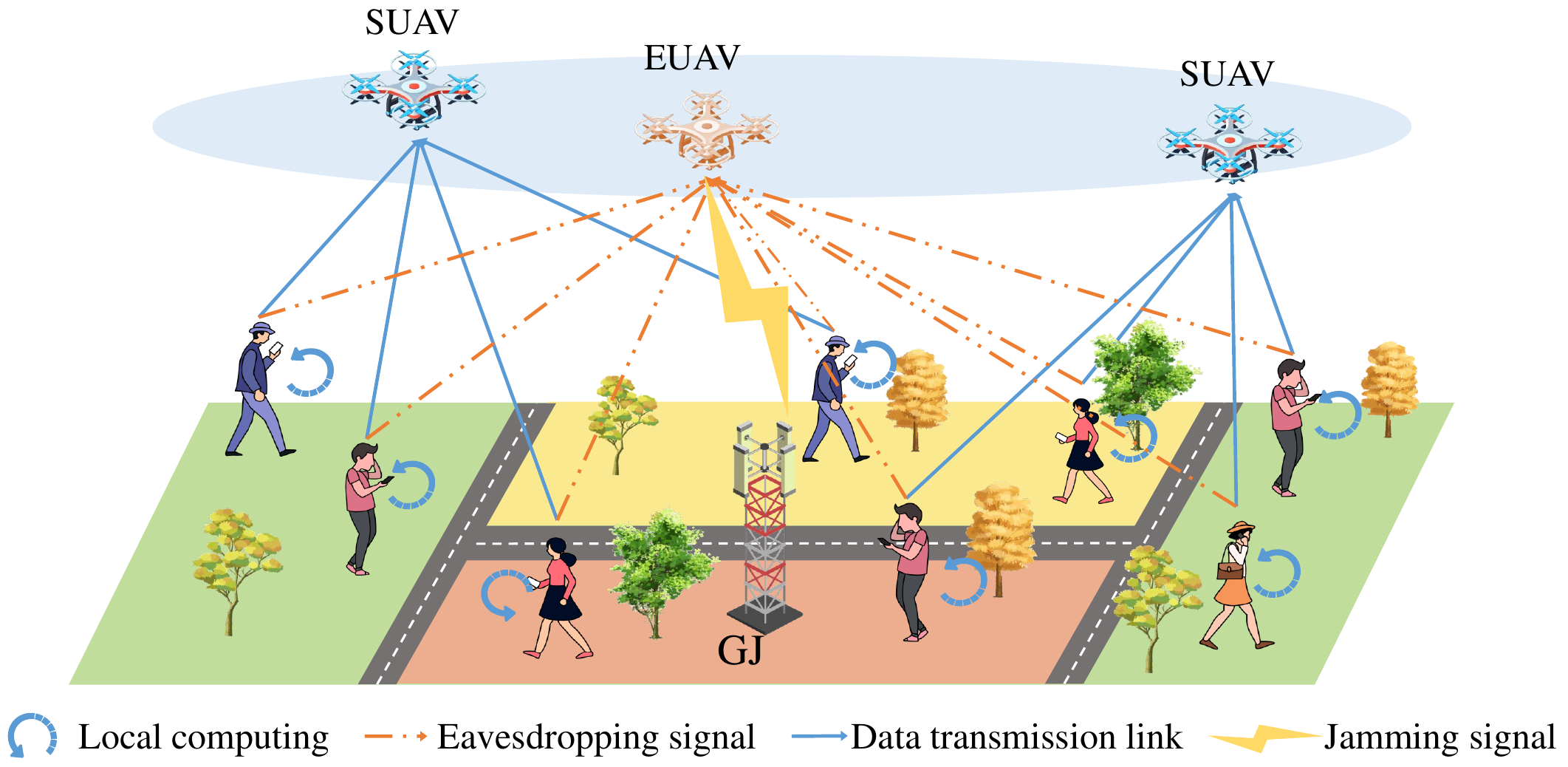}}
	\caption{A multiple UAVs enabled secure MEC scenario.}
	\label{Fig-1}
\end{figure}
As shown in Fig. \ref{Fig-1}, a secure MEC scenario consists of $I$ GUs denoted by $i\in\mathcal{I}=\left\{1,2,\cdots,I\right\}$, and $M$ SUAVs endowed with edge servers, in which $m\in\mathcal{M}=\left\{1,2,\cdots,M\right\}$. Due to the limitation of computing resources, GUs may be unable to complete their computation tasks within the required time. To satisfy the demand of quality of service (QoS), the MEC technology is applied, which enables GUs to offload their tasks to SUAVs partially for parallel processing. The binary variable $\lambda_{i,m}$ is deemed as an indicator for the connection relationship between GUs and SUAVs. Specifically, if GU $i$ connects to SUAV $m$, $\lambda_{i,m}=1$, and otherwise $\lambda_{i,m}=0$. A malicious EUAV flies randomly in the area and overhears the communication links. To ensure the security, a ground jammer (GJ) is deployed to inhibit overhearing from the EUAV and continues to broadcast the jamming signals. The total process is divided into $T$ time slots, each with a duration of $\tau$, denoted as $t\in\mathcal{T}=\left\{1,2,\cdots,T\right\}$. To characterize the three dimension coordinates of both GUs and UAVs, we employ the Cartesian coordinate system. Specifically, $w_i=\left(x_i,y_i\right)$ denotes the position of $i$-th GU. $m$-th SUAV is deployed at the horizon position $v_m=\left(x_m,y_m\right)$ with the flight height of $h_u$ and maintains its hovering throughout the entire time period. The EUAV flies across the area whose position message is previously known, and its position at time slot $t$ is $(x_e(t),y_e(t),h_e)$.

\subsection{Task Model}
To characterize the inherent attribute of heterogeneous tasks generated from GUs, we assume each GU has a computation task to be processed at time slot $t$, represented by $\left(L_i(t),c_i(t)\right)$. In detail, $L_i(t)$ is the data length of task $i$ and $c_i(t)$ is the corresponding task complexity, i.e., the required CPU cycles for calculating one bit data. Due to the limitation of practical applications, it is intractable to obtain the exact value of the task computation complexity \cite{Robust-Li}. To overcome the randomness, the task complexity is modeled as $c_i(t)=\bar{c}_i(t)+\Delta_i(t)$, in which $\bar{c}_i(t)$ is the estimated value of complexity, and $\Delta_i(t)$ is the random estimation error against the actual value. Although the exact value or distribution of the uncertain parameter $\Delta_i(t)$ is costly and unavailable, the first and second-order moment estimation information is easier to acquire via the historical statistical information \cite{Robust-Fan}. Thus, we assume the mean and variance value of $\Delta_i(t)$ measured from historical samples are knowable, denoted as $\mu_i(t)$ and $\sigma_i^2(t)$, respectively. Based on the statistical information, the uncertainty set for the random complexity estimation error is constructed as
\begin{equation}
	\mathcal{P}\triangleq \left\{ \mathbb{P}\in\mathcal{P}|\mathbb{E_P}(\Delta_i(t))=\mu_i(t), \mathbb{D_P}(\Delta_i(t))=\sigma_i^2(t)\right\},
\end{equation} 
where $\mathbb{P}$ is the possible probability distribution of $\Delta_i(t)$. $\mathbb{E_P}$ defines the expectation under distribution $\mathbb{P}$, and $\mathbb{D_P}$ defines the corresponding variance.

\subsection{Computation Model}
The tasks are assumed to be divisible, and the partial offloading mode is employed. We define the variable $\rho_i(t)\in\left[0,1\right]$ to represent the offloading ratio of GU $i$ at time slot $t$. Thus, the task can be divided into two parts: computed locally and offloaded to its corresponding SUAV.
\subsubsection{Local Computing}For the GUs with computing capabilities, they can process partial tasks locally. Let $f_g$ denote the CPU frequency of GUs for task processing. The computation delay for local computing is
\begin{equation}\label{e4}
	T_{i}^{l,c}(t)=\frac{(1-\rho_i(t))L_i(t)c_i(t)}{f_g}.
\end{equation}
We denote $\epsilon_g$ to represent the energy consumption coefficient for GUs related to the specific chip structure. Considering the random task computation complexity, the expected computing energy consumption for GU $i$ at time slot $t$ is
\begin{equation}
	\begin{split}
		E_{i}^{l,c}(t)=&\mathbb{E}\left\{\epsilon_g(1-\rho_i(t)c_i(t)L_i(t)f_g^2)\right\}\\
		=&\epsilon_g(1-\rho_i(t)(\bar{c}_i(t)+\mu_i(t))L_i(t)f_g^2).
	\end{split}
\end{equation}
\subsubsection{SUAV-based Edge Computing Model}The latency for SUAV-based edge computing model consists of transmission delay and computation delay. To facilitate the communication model, the orthogonal frequency division multiple access technique is applied  \cite{Multi-Agent-Ning}. We denote the bandwidth as $B_0$, and the power spectrum density of the additive white noise is represented as $n_0$. GUs transmit their tasks with the same power $p_u$. Besides, with the prior knowledge, SUAVs can distinguish the interference signals broadcasted from the GJ without being affected by the interfering information \cite{Secure-Lu}. Thus, the maximum achievable rate from GU $i$ to SUAV $m$ is
\begin{equation}
	r^{up}_{i,m}=B_0\log_2\left(1+\frac{p_ug_{i,m}}{n_0B_0}\right),
\end{equation}
where $g_{i,m}$ denotes the channel gain from GU $i$ to SUAV $m$, which can be characterized as a line of sight (LoS) link for a wide view \cite{Joint-Li}. Consequently, $g_{i,m}$ is calculated by
\begin{equation}
	g_{i,m}=\frac{g_0}{(x_i-x_m)^2+(y_i-y_m)^2+h_u^2},
\end{equation}
where $g_0$ is the channel gain at the reference distance $d_0=1$m. On the other hand, EUAV continues to overhear the communication link. However, they cannot distinguish the signals from the GJ, which is deemed as the noise. Let $p_j$ denote the interference signal power broadcasted from the GJ. The LoS channel model is employed and the eavesdropping rate from GU $i$ to the EUAV at time slot $t$ is
\begin{equation}
	r^{eav}_{i,e}(t)=B_0\log_2\left(1+\frac{p_ug_{i,e}(t)}{p_jg_{e,j}(t)+n_0B_0}\right).
\end{equation}
In detail,
\begin{equation}
	g_{i,e}(t)=\frac{g_0}{(x_i-x_e(t))^2+(y_i-y_e(t))^2+h_e^2},
\end{equation}
and
\begin{equation}
	g_{e,j}(t)=\frac{g_0}{(x_j-x_e(t))^2+(y_j-y_e(t))^2+h_e^2},
\end{equation}
are the channel gains from GU $i$ to the EUAV and from the GJ to the EUAV at time slot $t$, respectively. The maximum achievable secrecy transmission rate from GU $i$ to SUAV $m$ in time slot $t$ is defined as $r^{sec}_{i,m}(t)=\max\left\{r^{up}_{i,m}-r^{eav}_{i,e}(t),0\right\}$ \cite{Online-Ding}, and the transmission latency from GU $i$ to SUAV $m$ in time slot $t$ is calculated as
\begin{equation}
	T^{sec}_{i,m}(t)=\frac{\lambda_{i,m}\rho_i(t)L_i(t)}{r^{sec}_{i,m}(t)}.
\end{equation}
The transmission energy consumption from GU $i$ to SUAV $m$ at time slot $t$ is
\begin{equation}
	E_{i,m}^{e,d}(t)=p_uT^{sec}_{i,m}(t).
\end{equation}

Moreover, we assume the SUAVs allocate the same CPU frequency for the tasks they accommodate. Let $f_u$ denote the computational rate to process a single task, and the computation latency of task $i$ for edge computing model is \cite{Joint-You}
\begin{equation}
	T_{i}^{e,c}(t)=\frac{\rho_i(t)L_i(t)c_i(t)}{f_u}.
\end{equation}
Similarly, the expected computing energy consumption for edge computing to handle task $i$ is
\begin{equation}
	\begin{split}
		E_{i}^{e,c}(t)=&\mathbb{E}\left\{\epsilon_u\rho_i(t))c_i(t)L_i(t)f_u^2\right\}\\
		=&\epsilon_u\rho_i(t)(\bar{c}_i(t)+\mu_i(t))L_i(t)f_u^2,
	\end{split}
\end{equation}
in which $\epsilon_u$ is the energy cost coefficient of SUAVs.

Based on above analyses, the total latency including both transmission and computation delay is derived as
\begin{equation}\label{e3}
	T_{i}^{e}(t)=\sum_{m=1}^M T_{i,m}^{sec}(t)+T_{i}^{e,c}(t).
\end{equation}
Since the SUAVs hover throughout the time period, the energy cost for maintaining their flight is a constant value and can be ignored during the optimization. Consequently, let $\kappa$ represent the weighted variable to achieve a tradeoff for the energy consumption of SUAVs and GUs, and the total energy cost is expressed as
\begin{equation}
	\begin{split}
		E^{total}(t)=&\kappa\sum_{i=1}^I E^{l,c}_i(t)+(1-\kappa)\sum_{i=1}^I E^{e,c}_{i}(t)\\
		&+(1-\kappa)\sum_{i=1}^I\sum_{m=1}^M E^{e,d}_{i,m}(t).
	\end{split}
\end{equation}

\subsection{Problem Formulation}
Considering the limited battery capacities, we aim to minimize the total energy consumption of both GUs and SUAVs by optimizing the deployment of SUAVs $\bm{v}=\{v_m|\forall m \in \mathcal{M}\}$, the connection relationship of GUs and SUAVs $\bm{\lambda}=\{\lambda_{i,m}|\forall i\in\mathcal{I},m \in \mathcal{M}\}$, and the offloading ratio $\bm{\rho}=\{\rho_i(t)|\forall i \in \mathcal{I},\forall t \in \mathcal{T}\}$. To tackle the randomness of the uncertain parameter $\Delta_i(t)$, we formulate the chance constraints with restriction of the time latency. Mathematically, the original problem $\textbf{P0}$ is
\begin{subequations}
	\begin{align}
		\textbf{P0:}\quad &\min_{\bm{v},\bm{\lambda},\bm{\rho}} \sum_{t=1}^T E^{total}(t)\nonumber\\
		\textrm{s.t.} \quad &\mathbf{Pr}_{\mathbb{P}}\left\{T^{l,c}_{i}(t)\leq \tau \right\}\geq\alpha_i,\forall i\in \mathcal{I},\forall t \in \mathcal{T},\label{c1}\\
		\quad &\mathbf{Pr}_{\mathbb{P}}\left\{ T_{i}^{e}(t)\leq \tau \right\}\geq\alpha_i,\forall i\in \mathcal{I},\forall t \in \mathcal{T},\label{c2}\\
		\quad & v_m\in \biggl \{(x, y) \Bigg|\begin{aligned}X^{\min} \leq x \leq X^{\max}\\Y^{\min} \leq y \leq Y^{\max}\end{aligned} \biggl \},\forall m\in\mathcal{M},\label{c3}\\
		\quad &\sum_{m=1}^M\lambda_{i,m}=1,\label{c4}\\
		\quad &\rho_i(t) \in \left[ 0,1 \right],\forall i\in \mathcal{I},\forall t \in \mathcal{T}\label{c5},\\
		\quad &\lambda_{i,m}\in\{0,1\},\forall i\in \mathcal{I},\forall m\in \mathcal{M}\label{c6},
	\end{align}
\end{subequations}
where $\mathbf{Pr}_{\mathbb{P}}$ indicates that the chance constraint is formulated under the unknown distribution $\mathbb{P}$ for uncertain task computation complexity estimation error $\Delta_i(t)$. (\ref{c1}) and (\ref{c2}) are the chance constraints on $\mathbb{P}$, which indicate that the task should be completed within the time period under the safety factor $\alpha_i$. Regarding the random parameter $\Delta_i(t)$, the time latency for both local computing and SUAV based edge computing is not larger than $\tau$ at a possibility of $\alpha_i$. (\ref{c3}) is the deployment range limitations for SUAVs, in which $[X^{min},X^{max}]$ and $[Y^{min},Y^{max}]$ are deemed as the horizontal and vertical bounds of SUAVs, respectively. (\ref{c4}) is the access constraint. Wherein, each GU can only communicate with and offload its task to one SUAV  throughout the entire process. (\ref{c5}) is the range for $\rho_i(t)$, and (\ref{c6}) denotes $\lambda_{i,m}$ is a binary variable. However, without the distribution information for the uncertain task complexity, the problem $\textbf{P0}$ is obviously tricky to deal with.

\section{Algorithm Design\label{sec3}}
The total process can be divided into two phases. In the first phase, SUAVs are deployed at appropriate positions through the reasonable pre-deployment. In the second phase, GUs offload their data to the corresponding SUAV for offloading service. Consequently, in this section, we employ the K-means based algorithm and DRO based mechanism in an attempt to solve $\textbf{P0}$ with continuous variables and binary variables.

\subsection{Multi-SUAV Deployment}\label{ssec1}
The pre-deployment of multiple SUAVs and their connection relationships with GUs are crucial for enhancing system performance. A shorter distance between GUs and SUAVs can ultimately improve data transmission efficiency and contribute to less energy consumption. Consequently, the distance-driven K-means based algorithm is proposed in order to optimize the pre-deployment $\bm{v}$ of SUAVs by clustering GUs. After the SUAVs are deployed based on the clustering results, the connection relationships $\bm{\lambda}$ between GUs and SUAVs are established.

The process is provided in Algorithm \ref{alg1:K-means}. Firstly, with the pre-knowledge on the positions of GUs, we select $M$ points in the specific area as the initial deployment locations of SUAVs. Then, the distance between GU $i$ and SUAV $m$ is calculated as $d_{i,m}=\sqrt{(x_i-x_m)^2+(y_i-y_m)^2+h_u^2}$, and each GU is assigned to the nearest cluster. For each cluster $m$, the set of GUs in cluster $m$ is defined as $S_m$. The cluster center $v_m$ is updated as the average of the position coordinates of all GUs in the cluster $m$:
\begin{equation}\label{e18}
	v_m=\frac{\sum\limits_{k\in S_m}w_k}{|S_m|},
\end{equation}
where $|S_m|$ represents the number of GUs in cluster $m$. The above process is repeated until a convergence is achieved, and the obtained cluster centers are the final determined deployment positions of the SUAVs. 

\begin{algorithm}[t]
	\caption{K-means Based Multi-SUAV Deployment}\label{alg1:K-means}
	\begin{algorithmic}[1]
	\REQUIRE Locations of GUs.
	\STATE \textit{Initialization:} Set initial $\bm{v}$, $\lambda_{i,m}=0,\forall i\in\mathcal{I},\forall m\in\mathcal{M}$.
	\REPEAT
		\FOR{$m\in\mathcal{M}$}
			\STATE {Calculate distance $d_{i,m}$.\label{line-4}}
			\STATE {Assign GU $i$ to its nearest SUAV $m$, $\lambda_{i,m}=1$.\label{line-5}}
			\STATE Update $v_m$ based on (\ref{e18}).\label{line-6}
		\ENDFOR
	\UNTIL{the result converges.}
	\ENSURE SUAV deployment location $\bm{v}$ and the GU-SUAV connection $\bm{\lambda}$.
	\end{algorithmic}
\end{algorithm}
\subsection{Problem Transformation}\label{ssec2}
The determined deployment positions of SUAVs $\bm{v}$ and the connection relationships $\bm{\lambda}$ between GUs and SUAVs can serve as the basis for task offloading and data processing in the aerial MEC system. Then, the original problem $\textbf{P0}$ is further transformed into the following $\textbf{P1}$.
\begin{equation}\nonumber
	\begin{split}
		\textbf{P1:}\quad &\min_{\bm{\rho}} \sum_{t=1}^T E^{total}(t)\\
		\textrm{s.t.} \quad & (\ref{c1}),(\ref{c2}),(\ref{c5}).
	\end{split}
\end{equation}
However, the lack of the distribution information for the random parameter $\Delta_i(t)$ leads to the difficulty in handling the chance constraints (\ref{c1}) and (\ref{c2}). In response to this issue, by considering the worst-case scenario among all possible distributions, we aim to obtain a conservative and approximate estimation via the DRO method. In detail, let $\underset{\mathbb{P}\in\mathcal{P}}{inf}$ indicate the lower bound of possibility for all potential distributions under the uncertainty set $\mathcal{P}$. By employing the DRO method, the chance constraints (\ref{c1}) and (\ref{c2}) can be further transformed into DRCCs under the worst case of the distribution for random parameters, detailed as
\begin{equation}\label{drcc-1}
	\underset{\mathbb{P}\in\mathcal{P}}{inf}\quad\mathbf{Pr}_{\mathbb{P}}\left\{T^{e}_{i}(t)\leq \tau \right\}\geq\alpha_i,\forall i\in \mathcal{I},\forall t \in \mathcal{T},
\end{equation}
and
\begin{equation}\label{drcc-2}
	\underset{\mathbb{P}\in\mathcal{P}}{inf}\quad\mathbf{Pr}_{\mathbb{P}}\left\{(T^{l,c}_{i}(t)\leq \tau \right\}\geq\alpha_i,\forall i\in \mathcal{I},\forall t \in \mathcal{T},
\end{equation}
respectively, ensuring that the requirements for system performance can be satisfied even in the worst situations. However, note that DRCCs (\ref{drcc-1}) and (\ref{drcc-2}) embodying the uncertain parameter $\Delta_i(t)$ are still tricky to tackle.
\subsection{CVaR-based Optimization}\label{ssec3}
CVaR, known as a widely used tool in the field of risk assessment and optimization, plays a crucial role in dealing with uncertainty issues. Defined as the conditional expected loss exceeding the value at risk under the safety factor $\alpha$, CVaR further takes the tail loss information beyond the value at risk into account and enables a more comprehensive assessment of risk \cite{Optimization-Tyrrell, Distributionally-Jia}. Let $\underset{\mathbb{P}\in \mathcal{P}}{sup}$ denote the upper bound of possible distribution under the uncertainty set $\mathcal{P}$. For the loss function $\phi(\xi)=\Theta\xi+\theta^0$ regarding the random parameter $\xi$, the worst-case CVaR $\underset{\mathbb{P}\in\mathcal{P}}{sup}\quad\mathbb{P}-CVaR_{\alpha}(\phi(\xi))$ can be further converted into an SOCP \cite{Distributionally-Ding}, i.e.,
	\begin{equation}\label{e1}
		\begin{split}
			&\inf\quad \beta +\frac{1}{1-\alpha}\left( e+s \right) ,\\
			&e-\theta ^0+\beta +q-\Theta \mu -z>0,\\
			&e\geq 0, z> 0,\\
			&\begin{Vmatrix}
					q\\
					\Theta\sigma\\
					z-s
			\end{Vmatrix} \leq z+s,\\
		\end{split}
	\end{equation}
in which $\beta$, $e$, $q$, $z$, and $s$ are auxiliary variables. $\mu$ and $\sigma$ are the mean and standard deviation of random parameter $\xi$, respectively. Meanwhile, for the loss function $\phi(\xi)$, the DRCC be constituted by the worst-case CVaR constraint for a conservative estimation \cite{Distributionally-Zymler}, i.e., 
\begin{equation}\label{e2}
	\begin{split}
	\underset{\mathbb{P}\in \mathcal{P}}{inf}\quad \mathbb{P}\left\{\phi \left( \xi \right) \leq 0\right\}  \geq \alpha & \\
	\Leftrightarrow \underset{\mathbb{P}\in \mathcal{P}}{sup}\quad \mathbb{P}-CVaR_{\alpha}&\left( \phi \left( \xi \right) \right) \leq 0,\forall \mathbb{P}\in \mathcal{P}.
	\end{split}
\end{equation}
Thus, DRCCs (\ref{drcc-1}) and (\ref{drcc-2}) concerning random parameter $\Delta_i(t)$ can be derived in SOCP forms, detailed as
\begin{equation}\label{drcc-3}
	\begin{cases}
		&\inf\quad\beta_{i,t,1} +\frac{1}{1-\alpha_i}\left( e_{i,t,1}+s_{i,t,1}\right)\leq 0,\\
		&e_{i,t,1}-\theta_{i,1}(t)+\beta_{i,t,1}+q_{i,t,1}>\Theta_{i,1}(t)\mu_{i}(t)+z_{i,t,1},\\
		&e_{i,t,1}\geq 0,z_{i,t,1}> 0,\\
		&\begin{Vmatrix}
				q_{i,t,1}\\
				\Theta_{i,1}(t)\sigma_{i}(t)\\
				z_{i,t,1}-s_{i,t,1}
		\end{Vmatrix} \leq z_{i,t,1}+s_{i,t,1},\\
	\end{cases}
\end{equation}
and
\begin{equation}\label{drcc-4}
	\begin{cases}
		&\inf\quad\beta_{i,t,2} +\frac{1}{1-\alpha_i}\left( e_{i,t,2}+s_{i,t,2}\right)\leq 0,\\
		&e_{i,t,2}-\theta_{i,2}(t)+\beta_{i,t,2}+q_{i,t,2}>\Theta_{i,2}(t)\mu_{i}(t)+z_{i,t,2},\\
		&e_{i,t,2}\geq 0,z_{i,t,2}> 0,\\
		&\begin{Vmatrix}
				q_{i,t,2}\\
				\Theta_{i,2}(t)\sigma_{i}(t)\\
				z_{i,t,2}-s_{i,t,2}
		\end{Vmatrix} \leq z_{i,t,2}+s_{i,t,2},\\
	\end{cases}
\end{equation}
respectively, where $\Theta_{i,1}(t)=1-\rho_i(t)$ and $\Theta_{i,2}(t)=\rho_i(t)$. Besides,$\theta_{i,1}(t)=(1-\rho_i(t))\bar{c}_i(t)-\frac{\tau_i(t) f_g}{L_i(t)}$ and $\theta_{i,2}(t)=\rho_i(t)\bar{c}_i(t)+\sum\limits_{m=1}^M \frac{\lambda_{i,m}(t)\rho_i(t)f_u}{r^{sec}_{i,m}(t)}-\frac{\tau_i(t) f_u}{L_i(t)}$ are obtained according to the expression of $T_i^{l,c}(t)$ and $T_i^e(t)$ in equations (\ref{e4}) and (\ref{e3}), respectively. Consequently, by introducing the theory on CVaR, the DRCCs related to random parameters are transformed into the forms of convex SOCP. Besides, with the DRCC transformation and CVaR approximation, $\textbf{P1}$ is estimated into $\textbf{P2}$ with a conservative method,
\begin{equation}\nonumber
	\begin{split}
		\textbf{P2:}\quad &\min_{\bm{\rho},\bm{\beta},\bm{e},\bm{q},\bm{z},\bm{s}} \sum_{t=1}^T E^{total}(t)\\
		\textrm{s.t.} \quad & (\ref{c5}),(\ref{drcc-3}),(\ref{drcc-4}),
	\end{split}
\end{equation}
where $\bm{\beta},\bm{e},\bm{q},\bm{z},\bm{s}$ are the sets for all auxiliary variables. Note that $\textbf{P2}$ is a standard convex problem with SOCP constraints, we use CVX to solve it.
\subsection{Global Algorithm}\label{ssec4}
The global algorithm is detailed as follows. To solve $\textbf{P0}$, we first operate Algorithm \ref{alg1:K-means} to cluster GUs and deploy SUAVs at the center of each cluster. Then, by substituting the determined SUAV deployment locations $\bm{v}$ and the GU-SUAV connections $\bm{\lambda}$,  $\textbf{P0}$ is transformed into $\textbf{P1}$, which further turns into $\textbf{P2}$ via the DRO method and CVaR mechanism. Finally, $\textbf{P2}$ concerning the offloading ratios $\bm{\rho}$ is solved with CVX.

\section{Simulation Results\label{sec4}}

\begin{figure}[t]
	\centering{\includegraphics[scale=0.41]{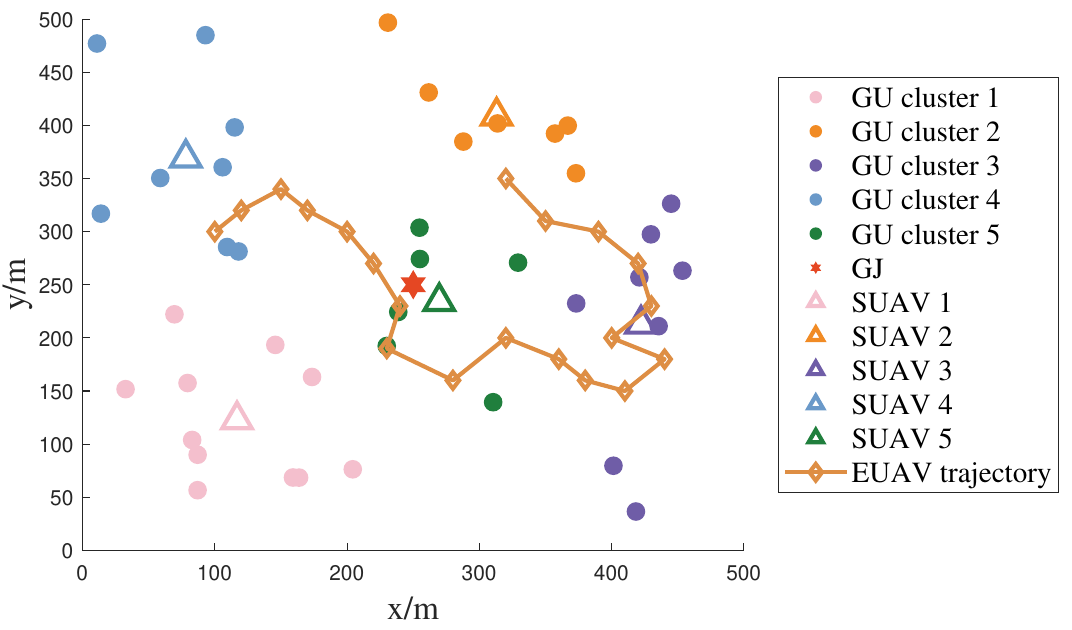}}
	\caption{Simulation scenario and deployment results (40 GUs and 5 SUAVs).}\label{Fig-2}
\end{figure}

\begin{figure}[t]
	\centering
	\subfloat[Total energy consumption.]{
		\includegraphics[scale=0.41]{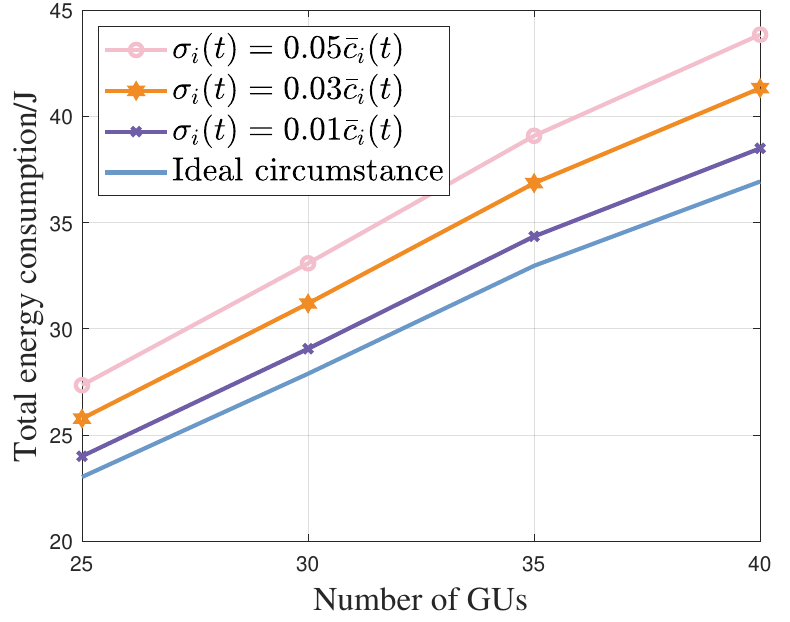}
	}
	\quad
	\subfloat[Offloaded data amount.]{
		\includegraphics[scale=0.41]{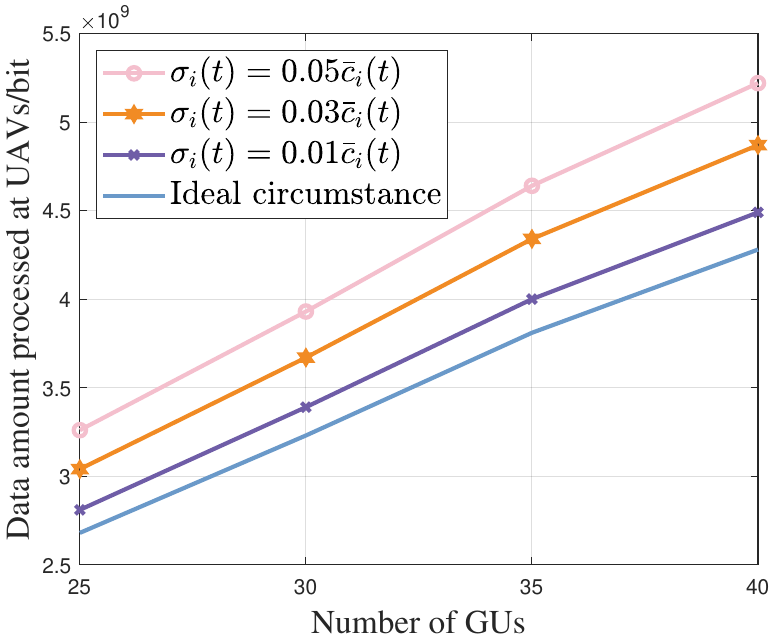}
	}
	\caption{Impact of random task computation complexity.}\label{Fig-3}
\end{figure}
In this section, we conduct numerical simulations to evaluate the performance of the proposed algorithms. We consider a $500\text{m}\times 500\text{m}$ area with 5 SUAVs and 20 time slots, with $\tau=5$s. The location of the GJ is set at $(250\text{m},250\text{m})$. The flight height for both SUAVs and the EUAV is 100m. The data length for GUs ranges within $[10,20]$Mbits and the task estimated complexity is $[10,100]$cycles/bit. $\mu_i(t)=0$cycles/bit. The allocated bandwidth for each GU is 10MHz. The transmission powers for GUs and the GJ are 2W and 20W, respectively. $g_0=-50$dB and $n_0=174$dBm/Hz. The computing capacities of GUs and SUAVs are $f_g=10^8$Hz and $f_u=10^9$Hz, respectively. $\epsilon_g$ and $\epsilon_u$ are set as $10^{-28}$. The weighted factor $\kappa$ is 0.2. A simulation example is shown in Fig. \ref{Fig-2}, including the distributions of 40 GUs, the flight trajectory of the EUAV, and the deployment of SUAVs obtained via Algorithm \ref{alg1:K-means}.

The effect of the potential task computation complexity estimation error considered in our scenario is discussed in Fig. \ref{Fig-3}. As $\sigma_i(t)$ (i.e., the standard deviation of random computation complexity) increases, the total energy consumption appears a growing trend, and GUs offload more data to the SUAVs. Due to the existence of the uncertain computational complexities, in order to meet the deadline requirements for task completion, GUs need to offload more data to the SUAVs, resulting in more energy consumption. Besides, more energy is consumed compared with the ideal circumstance, in which the task computation complexities are obtained precisely and the energy is minimized via non-robust traditional optimization methods. This trend proves the robustness of the proposed method.
\begin{figure}[t]
	\centering
	\subfloat[Impact of safety factor.]{
		\includegraphics[scale=0.4]{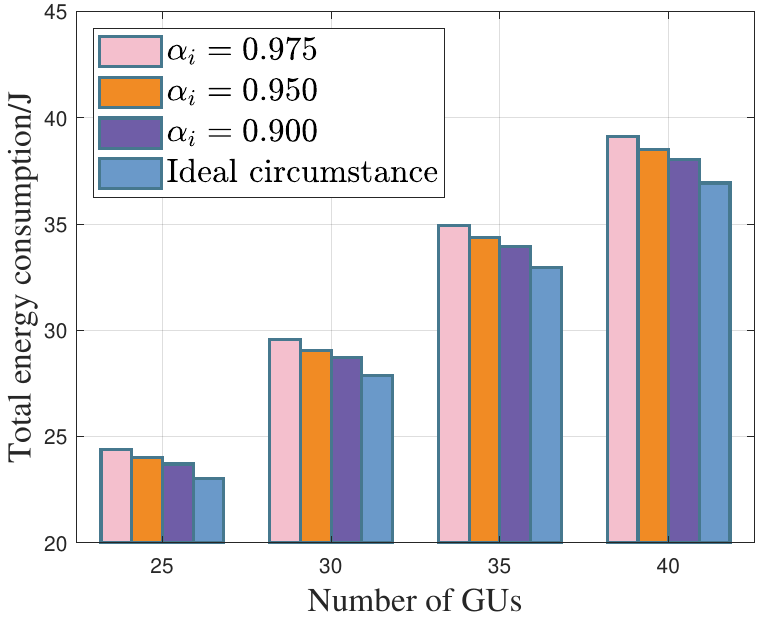}
	}
	\quad
	\subfloat[Impact of GU computing capacity.]{
		\includegraphics[scale=0.4]{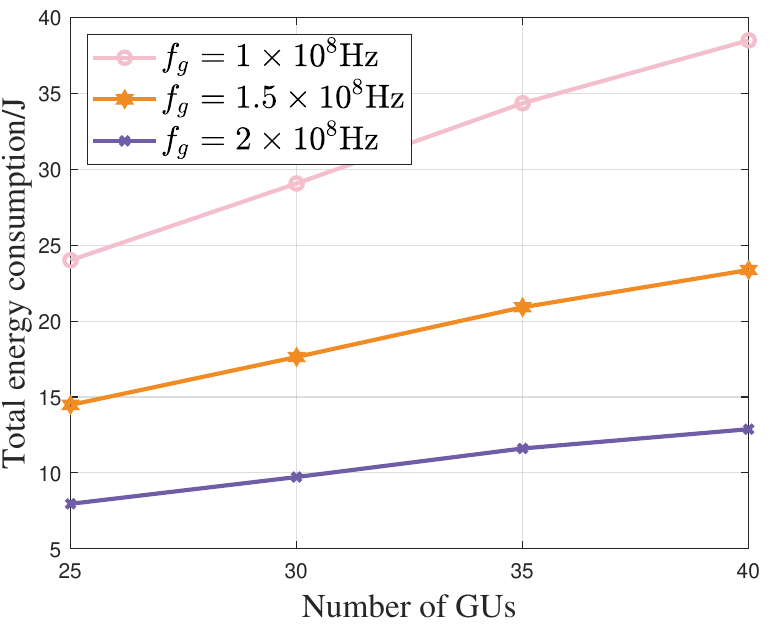}
	}
	\caption{Performance with different numbers of GUs.}\label{Fig-4}
\end{figure}

Furthermore, the impact of the safety factor on the system performance is analyzed in Fig. \ref{Fig-4}(a). It is observed that as $\alpha_i$ increases, more energy is consumed. This is because to ensure the more reliability for completing tasks, the offloading ratios are optimized more conservatively, which leads to an increment in energy consumption. Fig. \ref{Fig-4}(b) evaluates the influence of GU computing capacities on total energy cost. Note that the total energy consumption decreases with the increment of $f_g$. It is explained that due to the improvement of GU computing capacities, the tasks can be completed more efficiently locally, resulting in fewer tasks offloaded to SUAVs. In such case, the energy consumption for data transmission and SUAVs computation is decreased, leading to less energy consumption of the system.

\section{Conclusions\label{sec5}}
In this paper, we focused on the energy minimization for multiple UAVs enabled secure and robust MEC scenario considering the uncertain task computation complexities. We constructed an uncertainty set for the random parameters and formulated the problem with the chance constraints on the latency. Moreover, the K-means based algorithm was designed for the SUAV deployment and GU-SUAV connection. We further employed the DRO method for a conservative approximation and introduced the CVaR theory for convex optimization. Finally, the robustness of proposed algorithms was verified by simulation experiments, with about 5\% more energy consumption compared with the ideal circumstance. In addition, we also analyzed the impact of system parameters on the overall performance.
\bibliographystyle{IEEEtran}
\bibliography{reference.bib}
\end{document}